\documentclass[12pt]{article}\usepackage{graphicx,amsfonts,amsmath,amssymb,latexsym}
\usepackage{graphicx}

\renewcommand{\title}[1] {%
\begingroup\begin{center}\vspace{0.0cm}\bf\Large
\addtolength{\baselineskip}{1mm} #1 \end{center}\endgroup}

\renewcommand{\author}[1] {%
\begingroup\begin{center}\vspace{0.2cm}\bf #1 \vspace{0.2cm}
\end{center}\endgroup}

\newcommand{\address}[1] {%
\begingroup\begin{center} #1 \end{center}\endgroup}

\newcommand{\addressemail}[1] {%
\begingroup\begin{center}\vspace{3mm}
\vskip-\baselineskip #1 \end{center}\endgroup}

\newtheorem{thm}{Theorem}[section]
\newtheorem{defin}[thm]{Definition}

\newtheorem{prop}[thm]{Proposition}

\newcommand\Rb{\mathbb{R}}
\newcommand\Cb{\mathbb{C}}

\newcommand\Okr{\mathcal{O}}
\newcommand\rvac{|0\rangle}
\newcommand\lvac{\langle 0|}
\newcommand\ben{\begin{equation*}}
\newcommand\ebn{\end{equation*}}
\newcommand\be{\begin{equation}}
\newcommand\eb{\end{equation}}

\numberwithin{equation}{section}

\begin{document}
\title{Point interactions in one dimension \\
  and holonomic quantum fields}
 \author{ Oleg Lisovyy}
 \address{Bogolyubov Institute for Theoretical Physics \\
          14b Metrologichna str., 03143, Kyiv, Ukraine \\
          $\;$ \\
          School of Theoretical Physics, \\
          Dublin Institute for Advanced Studies, \\
          10 Burlington Road, Dublin 4, Ireland
}
 \addressemail{lisovyy-o@yandex.ru}
  \date{}

\begin{abstract}
We introduce and study a family of quantum fields, associated to
$\delta$-interactions in one dimension. These fields are analogous
to holonomic quantum fields of M.~Sato, T.~Miwa and M.~Jimbo.
Corresponding field operators belong to an infinite-dimensional
representation of the group $SL(2,\Rb)$ in the Fock space of
ordinary harmonic oscillator. We compute form factors of such
fields and their correlation functions, which are related to the
determinants of Schroedinger operators with a finite number of
point interactions. It is also shown that these determinants
coincide with tau functions, obtained through the trivialization
of the $\mathrm{det}^*$-bundle over a Grassmannian associated to a
family  of Schroedinger operators.
\end{abstract}

\section{Introduction}
The study of holonomic quantum fields \cite{smj} has led to
important advances in both integrable quantum field theory and
analytic theory of linear differential equations. In physical
language, these fields represent a particular case of the
canonical Bogolyubov transformations (see, for instance,
\cite{berezin}), which makes possible the exact computation of
their form factors and correlation functions.

The main physical examples of holonomic quantum fields are given
by the exponential fields of the sine-Gordon theory at the
free-fermion point ($\text{SG}_{ff}$) \cite{bernard,coleman,
schroer} and order/disorder variables of the two-dimensional Ising
model \cite{itzykson, kadanoff_ceva}. Both models possess an
underlying free-fermion structure. However, their observables are
nonlinear in terms of free fields (they are represented by ordered
exponentials of fermion bilinears) and have nontrivial braiding
relations with them. This nonlinearity leads to interesting
correlation functions.

Correlators of holonomic quantum fields are usually called tau
functions. In the special case of the $\text{SG}_{ff}$-theory they
have the meaning of determinants of Dirac operators with branching
points on the Euclidean plane. An attempt to give a geometric
definition of such tau functions was made in \cite{pacific}. It
was based on the approach, developed in the article \cite{cauchy},
where the tau function of the Schlesinger system was related to
the determinant of a singular Cauchy-Riemann operator. The idea of
\cite{pacific} was to consider a family $\mathcal{A}$ of Dirac
operators, parametrized by the coordinates of branch\-points, and
to associate to each of these operators a subspace of boundary
values of local solutions of the Dirac equation. These subspaces
are then embedded into an infinite-dimensional grassmannian. One
can construct \textit{\`a la} Segal-Wilson \cite{segal} the
$\mathrm{det}^*$-bundle over this grassmannian and its canonical
section~$\sigma$. Next, using the Green function of the Dirac
operator, one may endow the $\mathrm{det}^*$-bundle with another,
trivializing section. This latter allows to identify $\sigma$ with
a (tau) function on~$\mathcal{A}$.

The main goal of the present paper is to explain the concept of
holonomic quantum fields and the above definition of the tau
function with a simple example. It appears that these two notions
naturally emerge in the calculation of the resolvent of the
Schroedinger operator with $\delta$-interactions in one dimension.
Though such operators and their resolvents have already been
extensively discussed in physical and mathematical literature
(see, for example, the monograph \cite{albeverio} and references
therein), holonomic quantum fields allow to examine this fairly
classical subject from a new point of view.

Moreover, we believe that the ideology developed in this paper
applies as well to certain unsolved quantum mechanical problems,
the most appealing one being the computation of the fermionic
vacuum quantum numbers, induced by a finite number of magnetic
vortices in $2+1$ dimensions. In practice, the latter problem
reduces to the calculation of the resolvent of the Dirac
hamiltonian with point sources. At present, the answer is known
only in the case of a single vortex \cite{sitenko}. It seems,
however, that the multivortex resolvent can be obtained from the
correlation functions and form factors of certain Bogolyubov
transformations, generalizing exponential fields of the
$\text{SG}_{ff}$-theory.

This paper is organized as follows. After introducing basic
notations and terminology in Section~2, we turn in the next
section to the calculation of the resolvent of the Schroedinger
operator with $\delta$-interactions. It is expressed (by the
formula (\ref{resolvent2h})) through the ratio of correlation
functions of certain local fields in the 1D quantum field theory
of free massive real bosons. These correlation functions are
computed in the lagrangian approach by an auxiliary integration
method. They  comprise the fields of two types: the free ones and
interacting fields, associated to delta-sources. These latter
represent the simplest prototypes of holonomic quantum fields,
since the operators, corresponding to them in the hamiltonian
picture, realize certain canonical Bogolyubov transformations of
the Heisenberg algebra. In Section~4, we compute form factors and
correlation functions of the fields, which correspond to more
general Bogolyubov transformations and belong to an
infinite-dimensional representation of the group $SL(2,\Rb)$ in
the Fock space of harmonic oscillator. Section~5 is devoted to the
definition and calculation of the tau function. It is obtained
through the trivialization of the $\mathrm{det}^*$-bundle over a
finite-dimensional grassmannian of boundary conditions, associated
to a family of Schroedinger operators with point interactions.
Appendix~A contains several formulas of real and complex gaussian
integration. We conclude with a brief discussion of possible
generalizations and open questions. \vspace{0.2cm}\\
\textbf{Acknowledgements}. The author is grateful to A.~I.~Bugrij,
M.~Z.~Iorgov, V.~N.~Roubtsov, V.~M.~Shadura and Yu.~O.~Sitenko for
stimulating discussions and useful comments. \vspace{0cm}

 \section{Schroedinger operator without point interactions}
 Let us first consider Schroedinger operator without point
 interactions,
 \ben
 L=-\frac{d^2\;}{dx^2},
 \ebn
 acting on functions from the Sobolev space $H^2(\Rb)$ as the second derivative.
 Its resolvent $(L-E)^{-1}$ is given by the integral operator with the kernel
 \be\label{ge}
 G_E(x,y)=\frac{e^{-m|x-y|}}{2m},
 \eb
 which can be easily evaluated by Fourier transformation. Here we
 have introduced the notation $E=-m^2$ and assumed that
 $\mathrm{Re}\,m>0$.

 Note that for real $E<0$ the resolvent kernel $G_E(x,y)$ coincides by
 construction with the two-point correlation function in the
 relativistic euclidean quantum field theory of free massive
 real bosons in one dimension. In particular, if we define the action
 \be\label{action} S_0[\varphi]=\frac12\int\limits_{-\infty}^{\infty}dx
 \;\varphi(x)\left(-\frac{d^2\;}{dx^2}+m^2\right)\varphi(x),\eb
 then $G_E(x,y)$ can be formally written through the ratio of two
 functional integrals:
 \be\label{2pcf}
 G_E(x,y)=\langle\varphi(x)\varphi(y)\rangle=
 \frac{\int\mathcal{D}\varphi\;\varphi(x)\varphi(y)\;e^{-S_0[\varphi]}}
 {\int\mathcal{D}\varphi\;e^{-S_0[\varphi]}}.
 \eb
 The theory, described by the action (\ref{action}), admits two
 natural interpretations:
 \begin{itemize} \item
 If we interpret our single dimension as space, then the
 action (\ref{action}) coincides with the energy functional of the infinite string
 in a parabolic well. The integral
 in the denominator of (\ref{2pcf}) represents string partition function,
 and two-point correlator $\langle\varphi(x)\varphi(y)\rangle$
 is the  thermodynamic average of the product of transverse
 coordinates of two different points of the string.
 \item
 On the other hand, the action (\ref{action}) describes the
 dynamics of harmonic oscillator in imaginary time. In this
 setting, correlation function
 $\langle\varphi(x)\varphi(y)\rangle$ may be interpreted as the vacuum
 expectation value of the ordered product of coordinate operators at two
 different times.
 \end{itemize}

 In order to introduce several important notations, let us briefly recall
 the hamiltonian approach to oscillator dynamics. Fields $\varphi(x)$ and
 $\pi(x)=\frac{1}{i}\frac{d\varphi(x)}{dx\;}$ here become operators, obeying
 the commutation relation $[\hat{\varphi},\hat{\pi}]=i$. The hamiltonian,
 being expressed in terms of $\hat{\varphi}$ and $\hat{\pi}$, has
 the form
 \ben
 \hat{H}=\frac12\left(\hat{\pi}^2+m^2\hat{\varphi}^2\right)-\frac{m}{2},
 \ebn
 where the constant term is subtracted for future convenience.
 Imaginary time evolution of an arbitrary operator $\hat{\Okr}$ is given by the equation
 \be\label{evolution}
 \hat{\Okr}(x)=e^{-\hat{H}x}\hat{\Okr}(0)\; e^{\hat{H}x}.
 \eb
 It is customary to define the creation-annihilation operators
 \ben
 a=\sqrt{\frac{m}{2}}\left(\hat{\varphi}(0)+\frac{i\,\hat{\pi}(0)}{m}\right),\qquad
 a^{\dag}=\sqrt{\frac{m}{2}}\left(\hat{\varphi}(0)-\frac{i\,\hat{\pi}(0)}{m}\right),
 \ebn
 satisfying canonical commutation relation $[a,a^{\dag}]=1$. The
 hamiltonian can then be rewritten in terms of these operators as
 $\hat{H}=m\,a^{\dag}a$. Vacuum vector $\rvac$ is fixed by conditions $a\rvac=0$, $\lvac
 {0\rangle}=1$. The operators that we wish to consider act in the Fock
 space $\mathcal{F}$, spanned by the orthonormal vectors
 \be\label{fock}
 |k\rangle=\frac{\left(a^{\dag}\right)^k}{\sqrt{k!}\;}\,\rvac,\qquad
 k=0,1,2\ldots,
 \eb
 constituting the set of hamiltonian eigenstates:
 $\hat{H}|k\rangle=km\,|k\rangle$.

 The computation of correlation functions of local fields in the hamiltonian approach is equivalent
 to the calculation of form factors, i.~e. matrix elements of the corresponding field operators
 in the orthonormal basis of eigenstates of $\hat{H}$. For
 instance, form factor expansion of the two-point correlator $\langle
 \Okr_1(x_1)\Okr_2(x_2)\rangle$ is written as
 \ben
 \langle \Okr_1(x_1)\Okr_2(x_2)\rangle=
 \lvac\hat{\Okr}_1(x_1)\hat{\Okr}_2(x_2)\rvac=
 \lvac\hat{\Okr}_1(0)\; e^{-\hat{H}(x_2-x_1)}\hat{\Okr}_2(0)\rvac=
 \ebn
 \be\label{2pffe}
 =\sum\limits_{k=0}^{\infty}\;\lvac\,\hat{\Okr}_1(0)|k\rangle\,
 \langle k|\hat{\Okr}_2(0)\rvac\;e^{-km(x_2-x_1)},
 \eb
 where we have assumed that $x_2\geq x_1$. One can obtain
 analogous expressions for the multipoint correlation functions,
 simply using the formula (\ref{evolution}), and inserting the appropriate number of times
 the resolution of the identity operator, $\mathbf{1}=\sum\limits_{k=0}^{\infty}|k\rangle\,\langle k|$,
 into the correlators.\vspace{0.2cm}\\
 \textbf{Example}. It is easiest to see how this scheme works on
 the example of the two-point correlation function
 $\langle\varphi(x)\varphi(y)\rangle$. Note that since the
 operators $\hat{\varphi}$ and $\hat{\pi}$ are given by
 \ben
 \hat{\varphi}(0)=\frac{1}{\sqrt{2m}}\left(a^{\dag}+a\right),\qquad
 \hat{\pi}(0)=i\,\sqrt{\frac{m}{2}}\left(a^{\dag}-a\right),
 \ebn
 the only non-zero form factors are
 \begin{eqnarray*}
 \langle k+1|\,\hat{\varphi}(0)|k\rangle &=&
 \langle k|\,\hat{\varphi}(0)|k+1\rangle=\sqrt{\frac{k+1}{2m}}\;,\\
 \langle k+1|\,\hat{\pi}(0)|k\rangle &=&
 -\langle k|\,\hat{\pi}(0)|k+1\rangle=i\,\sqrt{\frac{k+1}{2m}}\;,
 \end{eqnarray*}
 that is, the numbers of
 \textit{in}- and \textit{out}-particles should differ by 1.
 Therefore, in the expansion over intermediate states (\ref{2pffe}) for
 $\langle\varphi(x)\varphi(y)\rangle$ only the terms with $k=1$
 will remain, reproducing thus the formula (\ref{ge}).

 \section{Introducing $\delta$-interactions}
 Let us now see what happens if instead of $L$ we consider
 Schroedinger operator with a finite number of
 $\delta$-interactions,
 \ben
 L_{a,V}=-\frac{d^2\;}{dx^2}+V(x),\qquad\qquad V(x)=\sum\limits_{i=1}^N
 V_{i}\,\delta(x-a_i)\,.
 \ebn
 The most known way of calculating the resolvent $(L_{a,V}-E)^{-1}$
 is to expand it formally in a series in powers of $V$. Summing up this
 series, one obtains a compact expression for the resolvent kernel~$G_{E,\,V}(x,y)$:
 \be\label{resolvent1}
 G_{E,\,V}(x,y)=G_E(x,y)-\sum\limits_{i,j=1}^N
 G_E(x,a_i)\,U^{-1}_{ij}\,G_E(a_j,x),
 \eb
 Here $G_E(x,y)$ denotes the unperturbed resolvent (\ref{ge}) and the matrix $U$ is defined as
 \ben
 U_{ij}=\frac{1}{V_i}\;\delta_{ij}+G_{E}(a_i,a_j),\qquad\qquad
 i,j=1,\ldots,N.
 \ebn

 An alternative simple proof of this result follows from
 field-theoretic considerations. Again, for real negative $E=-m^2$
 the resolvent $G_{E,\,V}(x,y)$ coincides with the two-point correlator
 in the one-dimensional quantum field theory, described by the action
 \be\label{action2} S_V[\varphi]=\frac12\int\limits_{-\infty}^{\infty}dx
 \;\varphi(x)\left(-\frac{d^2\;}{dx^2}+m^2+V(x)\right)\varphi(x)=
 S_0[\varphi]+S_d[\varphi],\eb
 where
 \ben S_d[\varphi]=\frac12 \sum\limits_{i=1}^{N}V_i\,{\varphi^2(a_i)}.\ebn
 Thus we have
 \ben
 G_{E,\,V}(x,y)=\langle\varphi(x)\varphi(y)\rangle_V=
 \frac{\int\mathcal{D}\varphi\;\varphi(x)\varphi(y)\;e^{-S_V[\varphi]}}
 {\int\mathcal{D}\varphi\;e^{-S_V[\varphi]}}.
 \ebn
 The action (\ref{action2}) reproduces the energy of a one-dimensional string with $N$ masses
 attached to the points $a_1,\ldots,a_N$. We will assume for
 definiteness that
 these masses are all positive.

 The idea is to represent the factor $e^{-S_d[\varphi]}$,
 appearing in the functional integrals, as a gaussian integral
 over $N$ auxiliary variables $\mu_1,\ldots,\mu_N$:
 \ben
 e^{-S_d[\varphi]}=\frac{1\;\;}{\left(2\pi\right)^{N/2}\prod\nolimits_{j=1}^N
 \sqrt{V_j}}\;
 \int\limits_{-\infty}^{\infty}\ldots \int\limits_{-\infty}^{\infty}d\mu_1\ldots d\mu_N\;
 \exp\left\{-\sum\limits_{j=1}^N \mu_j^{\,2}/V_j+\sum\limits_{j=1}^N i\mu_j\,\varphi(a_j)\right\}.
 \ebn
 Now one can make use of the formula (\ref{krein1}) of the Appendix.
 The matrix $A$ in our case has continuous indices, and its
 inverse $A^{-1}$ should be replaced by the resolvent
 $(L+m^2)^{-1}$. The matrix $B$ is diagonal,
 \ben
 B_{kl}=\frac{1}{V_k}\;\delta_{kl},\qquad k,l=1,\ldots,N,
 \ebn
 and the matrix $C$ has one continuous and one discrete index:
 \ben
 C_{x,k}=-i\,\delta(x-a_k), \qquad k=1,\ldots,N,\quad x\in\Rb.
 \ebn
 Substituting these matrices into the relation (\ref{krein1}), we immediately obtain
 the formula (\ref{resolvent1}) for the resolvent kernel
 $G_{E,V}(x,y)$.\vspace{0.2cm}\\
 \textbf{Remark}. Schroedinger operators with
 $\delta$-interactions at the points $a_1,\ldots,a_N$ are
 rigorously defined as the elements of an $N$-parametric family of
 self-adjoint extensions of the operator
 $\dot{L}_a=-\frac{d^2\;}{dx^2}$ with the domain
 \ben
 \mathrm{dom}\,\dot{L}_a=\left\{f\in H^2(\Rb)\,|\;
 f(a_j)=0,\;j=1,\ldots,N\right\}.
 \ebn
 This operator has deficiency indices $(N,N)$ and thus the whole
 family of its self-adjoint extensions depends on $N^2$ parameters that can be incorporated into
 a hermitian $N\times N$ matrix $B$. Similarly to the above, one
 can associate to a general self-adjoint extension $\dot{L}_{a,B}$
 the action
 \ben
 S_B[\varphi,\bar{\varphi}]=\int\limits_{-\infty}^{\infty}dx\;\bar{\varphi}(x)(L+m^2)\varphi(x)+
 \sum\limits_{i,j=1}^{N}\bar{\varphi}(a_i)B_{ij}^{\,-1}\varphi(a_j).
 \ebn
 The kernel $G_{E,B}(x,y)$ of the resolvent
 $(\dot{L}_{a,B}+m^2)^{-1}$ coincides with the two-point
 correlation function
 \ben
 G_{E,B}(x,y)=\langle\varphi(x)\bar{\varphi}(y)\rangle_B=
 \frac{\int\mathcal{D}\varphi\,\mathcal{D}\bar{\varphi}\;\varphi(x)\bar{\varphi}(y)\;e^{-S_B[\varphi,\bar{\varphi}]}}
 {\int\mathcal{D}\varphi\,\mathcal{D}\bar{\varphi}\;e^{-S_B[\varphi,\bar{\varphi}]}}.
 \ebn
 It can be easily evaluated, using the above method of auxiliary
 fields and applying the formulas of complex gaussian integration from the Appendix.
 The result has the form (\ref{resolvent1}); the only
 difference is that the matrix $U$ now depends on $B$:
 \ben
 U_{ij}=B_{ij}+G_E(a_i,a_j).
 \ebn
 In what follows, however, we will be dealing only with local
 point interactions.
 \vspace{0.2cm}

 Let us consider yet another approach to the calculation of the
 resolvent $G_{E,V}(x,y)$. On the one hand, it is equal to pair correlation function
 in the theory with the action (\ref{action2}). However, it
 may also be expressed through
 the ratio of certain multipoint correlators in the theory, described by the unperturbed action
 (\ref{action}). Namely, one can write
 \be\label{resolvent2h}
 G_{E,V}(x,y)=\langle\varphi(x)\varphi(y)\rangle_V=
 \frac{\langle\Okr_{V_1}(a_1)\ldots\Okr_{V_N}(a_N)\varphi(x)\varphi(y)\rangle}
 {\langle\Okr_{V_1}(a_1)\ldots\Okr_{V_N}(a_N)\rangle}\, ,
 \eb
 where the local fields $\Okr_{V}(a)$ are defined as
 \be\label{expl}
 \Okr_{V}(a)=\exp\left\{-\frac12\, V\varphi^2(a)\right\}.
 \eb

 It should be pointed out that the correlation function
 $\langle\Okr_{V_1}(a_1)\ldots\Okr_{V_N}(a_N)\rangle$, standing in
 the denominator of (\ref{resolvent2h}), is equal to the ratio of
 partition functions of the string with and without attached point
 masses. It can be formally expressed through the determinants of
 Schroedinger operators:
 \be\label{connection}
 \langle\Okr_{V_1}(a_1)\ldots\Okr_{V_N}(a_N)\rangle=
 \sqrt{\frac{\mathrm{det}(L-E)}{\mathrm{det}(L_{a,V}-E)}}\,.
 \eb
 Repeating the trick with auxiliary fields and using the formula
 (\ref{ff}) of the Appendix, one may also compute this correlator
 explicitly:
 \be\label{korr}
 \langle\Okr_{V_1}(a_1)\ldots\Okr_{V_N}(a_N)\rangle=
 \biggl(\mathrm{det}\left\|\,\delta_{ij}+\sqrt{V_i V_j}\,G_{E}(a_i,a_j)\,
 \right\|\biggr)^{-1/2}\!\!\!,\qquad\quad i,j=1,\ldots,N.
 \eb

 Consider now the fields $\Okr_V$ in the hamiltonian picture.
 The operator $\hat{\Okr}_{V}=\exp\left\{-\frac12\,
 V\hat{\varphi}^{\,2}\right\}$ has remarkable equal-time
 commutation relations with the operators $\hat{\varphi}$ and $\hat{\pi}$:
 \begin{eqnarray}
 \hat{\Okr}_{V}(0)\hat{\varphi}(0)-\hat{\varphi}(0)\hat{\Okr}_{V}(0)&=&0,\label{comm1}\\
 \hat{\Okr}_{V}(0)\hat{\pi}(0)-\hat{\pi}(0)\hat{\Okr}_{V}(0)&=&-iV\hat{\varphi}(0)\hat{\Okr}_{V}(0).\label{comm2}
 \end{eqnarray}
 This is a manifestation of the fact that any correlator $f(x)=\langle\ldots
 \Okr_{V}(a)\varphi(x)\ldots\rangle$, being considered as a
 function of $x$, is a local solution of the Schroedinger equation $(L-E)f=0$,
 satisfying at the point $a$ the boundary condition
 \ben
 f(a+0)-f(a-0)=0,\qquad f'(a+0)-f'(a-0)=Vf(a).
 \ebn
 One can also rewrite the commutation relations (\ref{comm1})--(\ref{comm2}) in terms of the
 creation-annihilation operators:
 \be\label{bog1}
 \hat{\Okr}_{V}(0)\left(\begin{array}{l} a \\ a^{\dag} \end{array}
 \right)\hat{\Okr}^{-1}_{V}(0)=
 \left(\begin{array}{cc} 1+\frac{V}{2m} & \frac{V}{2m}\; \\ -\frac{V}{2m}\quad & 1-\frac{V}{2m} \end{array}\right)
 \left(\begin{array}{l} a \\ a^{\dag} \end{array}\right).
 \eb
 Therefore, $\hat{\Okr}_V$ realizes a Bogolyubov
 transformation, i.~e. linear transformation of the Heisenberg
 algebra, preserving canonical commutation relation
 $[a,a^{\dag}]=1$. The formula (\ref{bog1}) determines
 the operator $\hat{\Okr}_V$ almost completely. More precisely, $\hat{\Okr}_V$
 is fixed by (\ref{bog1}) up to a constant numerical factor.

 In the next section, we will compute form factors and correlation
 functions of the fields, correspon\-ding to more general Bogolyubov
 transformations. It should be emphasized that we will use only
 the relations of type (\ref{bog1}) and no reference will be made
 to the explicit formula (\ref{expl}). The reason for doing so is that it
 is not always clear, which operator should be associated to a
 given field and vice versa. This is the case, for instance, in the analysis of
 $\delta'$-interactions. Even more severe difficulties arise in
 some two-dimensional problems: the fields, realizing relevant
 Bogolyubov transformations, being themselves local, are not mutually local
 with the free fields. The most known example of this kind is given by the exponential fields
 in the $\text{SG}_{ff}$-theory.

 \section{Form factors and correlation functions \\ of Bogolyubov fields}
 Consider a linear transformation of the creation-annihilation
 operators
 \be\label{lmap}
 \Lambda:\;\left(\begin{array}{l} a \\ a^{\dag} \end{array}\right)
 \mapsto\left(\begin{array}{l} b \\ b^{\dag} \end{array}\right)=
 \left(\begin{array}{cc} \alpha & \beta \\ \gamma & \delta \end{array}\right)
 \left(\begin{array}{l} a \\ a^{\dag} \end{array}\right).
 \eb
 It preserves canonical commutation relation iff the real parameters
 $\alpha$, $\beta$, $\gamma$, $\delta$ satisfy the condition
 $\alpha\delta-\beta\gamma=1$. In the following, this condition is assumed to
 hold. We want
 to represent $\Lambda$ as a similarity transformation. Namely,
 we are looking for the invertible operator
 $\hat{\Okr}_{\Lambda}$,
 such that
 \be\label{boggen}
 \hat{\Okr}_{\Lambda}\left(\begin{array}{l} a \\ a^{\dag} \end{array}\right)\hat{\Okr}_{\Lambda}^{-1}=
 \Lambda\left(\begin{array}{l} a \\ a^{\dag} \end{array}\right).
 \eb
 These operators are called Bogolyubov transformations, and the
 corresponding local fields will be called Bogolyubov fields in
 the rest of this paper\footnote{Actually, there already exist several names for some special fields of
 this type, to mention only `holonomic quantum fields', `exponential fields' and `monodromy fields'.
 All these names, however, do not reflect common structure of such fields,
 and are related to the particularities of different two-dimensional problems.
 That is why we believe that the name `Bogolyubov fields'
 would be more appropriate.}. The operators $\hat{\Okr}_{\Lambda}$ are
 determined by (\ref{boggen}) up to a constant, realizing thus an
 infinite-dimensional projective representation of the group $SL(2,\Rb)$
 in the Fock space of our harmonic oscillator.

 In order to construct $\hat{\Okr}_{\Lambda}$ in terms of $a$ and
 $a^{\dag}$, one should examine the properties of basic Bogolyubov transformations:
 \ben
 \hat{P}_{\lambda}=e^{\,\frac12\,\lambda\left(a^{\dag}\right)^2},\qquad
 \hat{R}_{\nu}=e^{\,\frac12\,\nu \,a^2},
 \ebn
 \ben
 \hat{Q}_{\mu}=\; :e^{\,\mu \,a^{\dag}a}:\; \substack{\text{\textit{def}}\\=\\ \;}
 \;\sum\limits_{n=0}^{\infty}\,\frac{\mu^n}{n!}\,\left(a^{\dag}\right)^n a^n.
 \ebn
 It is easy to check that induced linear transformations have
 the form
 \begin{eqnarray}
 \hat{P}_{\lambda}\left(\begin{array}{l} a \\ a^{\dag}
 \end{array}\right)\hat{P}_{\lambda}^{-1} &=&
 \left(\begin{array}{cc} 1 & -\lambda \\ 0 & \; 1 \end{array}\right)
 \left(\begin{array}{l} a \\ a^{\dag} \end{array}\right), \label{comp}\\
 \hat{Q}_{\mu}\left(\begin{array}{l} a \\ a^{\dag}
 \end{array}\right)\hat{Q}_{\mu}^{-1} &=&
 \left(\begin{array}{cc} \frac{1}{1+\mu} & 0 \\ 0 &  1+\mu \end{array}\right)
 \left(\begin{array}{l} a \\ a^{\dag} \end{array}\right), \label{comq}\\
 \hat{R}_{\nu}\left(\begin{array}{l} a \\ a^{\dag}
 \end{array}\right)\hat{R}_{\nu}^{-1} &=&
 \left(\begin{array}{cc} 1 & 0 \\ \nu &  1 \end{array}\right)
 \left(\begin{array}{l} a \\ a^{\dag}
 \end{array}\right).\label{comr}
 \end{eqnarray}
 General Bogolyubov transformation (\ref{boggen}) is then given by
 \be\label{btr1}
 \hat{\Okr}_{\lambda,\mu,\nu}=\hat{P}_{\lambda}\hat{Q}_{\mu}\hat{R}_{\nu}=
 \; :\exp\left\{\frac12\,\lambda\left(a^{\dag}\right)^2+\mu \,a^{\dag}a+\frac12\,\nu\, a^2\right\} :
 \eb
 Parameters $\alpha$, $\beta$, $\gamma$, $\delta$ of the
 corresponding linear map
 $\Lambda=\Lambda_{R_{\nu}}\Lambda_{Q_{\mu}}\Lambda_{P_{\lambda}}$
 are related to $\lambda$, $\mu$ and $\nu$ by the following formulas:
 \be\label{abcd}
 \alpha=\frac{1}{1+\mu},\qquad \beta=-\frac{\lambda}{1+\mu},\qquad
 \gamma=\frac{\nu}{1+\mu},\qquad \delta=1+\mu-\frac{\lambda\,\nu}{1+\mu}\,.
 \eb
 Note that writing $\hat{\Okr}_{\Lambda}$ in the form (\ref{btr1}), we adopt
 the convention $\langle\Okr_{\Lambda}\rangle\;
 \substack{\text{\textit{def}} \\ = \\ \;}
 \;\langle 0|\hat{\Okr}_{\Lambda}|0\rangle=1$. This normalization will be used hereafter.
 Together with the relation (\ref{boggen}), it completely fixes the
 operator~$\hat{\Okr}_{\Lambda}$.\vspace{0.3cm}\\
 \textbf{Remark}. One-point function of the Bogolyubov field  $\Okr_V$, associated to
 a $\delta$-interaction, is \textit{not} equal to $1$. It may be
 determined from the formula (\ref{korr}) by setting $N=1$:
 \be\label{1point}
 \langle\Okr_V\rangle=\left(1+\frac{V}{2m}\right)^{-1/2}.
 \eb
 It can also be computed in the hamiltonian approach, being
 rewritten as the vacuum expectation value
 \be\label{aux1}
 \langle\Okr_V\rangle=\langle0|\hat{\Okr}_V|0\rangle=
 \langle0|e^{-\frac12\,V\hat{\varphi}^2}|0\rangle=
 \sum\limits_{n=0}^{\infty}\frac{1}{n!}\left(-\frac{V}{4m}\right)^n
 \langle0|\left(a+a^{\dag}\right)^{2n}|0\rangle.
 \eb
 Using Wick's theorem, one obtains
 $\langle0|\left(a+a^{\dag}\right)^{2n}|0\rangle=(2n-1)!!$. Now
 remark that the Taylor expansion of the function
 \be\label{series}
 (1+x)^{-1/2}=1+\sum\limits_{n=1}^{\infty}\frac{(2n-1)!!}{n!}\left(-\frac{x}{2}\right)^n
 \eb
 coincides with (\ref{aux1}), if we set $x=\frac{V}{2m}$. Thus
 we recover the formula (\ref{1point}).

 Parameters $\lambda$, $\mu$ and $\nu$, which correspond to the field $\Okr_V$, are
 determined from the comparison of (\ref{bog1}) and (\ref{abcd}).
 The result is
 \be\label{lmn}
 \lambda=\mu=\nu=-\frac{V/2m}{1+V/2m}.
 \eb
 This leads to the following representation of the operator
 $\hat{\Okr}_V$:
 \be\label{pifield1}
 \hat{\Okr}_V=\left(1+\frac{V}{2m}\right)^{-1/2}
 :\;\exp\left\{-\frac12\,\frac{V/2m}{1+V/2m}\,\left(a+a^{\dag}\right)^2\right\}:
 \eb
 \vspace{0.3cm}

 We now turn to the calculation of form factors
 $\langle k|\hat{\Okr}_{\lambda,\mu,\nu}|l\rangle$ of the
 operator (\ref{btr1}). It is clear that such form factors will be
 non-zero only if the numbers of particles in \textit{in}- and
 \textit{out}-state are simultaneously even or odd.
 It is convenient to use instead of $\langle k|\hat{\Okr}_{\lambda,\mu,\nu}|l\rangle$
 the auxiliary variables
 \ben
 F_{k,l}=\sqrt{k!}\,\sqrt{l!}\;\langle
 k|\hat{\Okr}_{\lambda,\mu,\nu}|l\rangle=
 \langle 0|a^k\,\hat{P}_{\lambda}\hat{Q}_{\mu}\hat{R}_{\nu}
 \left(a^{\dag}\right)^l|0\rangle.
 \ebn
 Using the relations (\ref{comp})--(\ref{comr}) in the last expression, one can pull
 the operator $\hat{P}_{\lambda}$  through $a^k$, and the
 operators $\hat{Q}_{\mu}$, $\hat{R}_{\nu}$ through $\left(a^{\dag}\right)^l$:
 \ben
 F_{k,l}=\langle 0|\hat{P}_{\lambda}\left(a+\lambda\, a^{\dag}\right)^k\,
 \hat{Q}_{\mu}\hat{R}_{\nu} \left(a^{\dag}\right)^l|0\rangle=
 \langle 0|\hat{P}_{\lambda}\left(a+\lambda\, a^{\dag}\right)^k\,
 \hat{Q}_{\mu}
 \left(\nu\,a+a^{\dag}\right)^l\hat{R}_{\nu}|0\rangle=
 \ebn
 \ben
 =\langle 0|\hat{P}_{\lambda}\left(a+\lambda\, a^{\dag}\right)^k\,
 \Bigl(\nu(1+\mu)^{-1}a+(1+\mu)\,a^{\dag}\Bigr)^l
 \hat{Q}_{\mu}\hat{R}_{\nu}|0\rangle.
 \ebn
 Next, since we have
 \ben
 \langle 0|\hat{P}_{\lambda}=\langle
 0|\hat{Q}_{\mu}=\langle0|,\qquad\qquad
 \hat{Q}_{\mu}|0\rangle=\hat{R}_{\nu}|0\rangle=|0\rangle,
 \ebn
 the variable $F_{k,l}$ may be rewritten as the vacuum expectation
 value of the product of certain linear combinations of the creation-annihilation operators:
 \ben
 F_{k,l}=\langle 0|\left(a+\lambda\, a^{\dag}\right)^k\,
 \Bigl(\nu(1+\mu)^{-1}a+(1+\mu)\,a^{\dag}\Bigr)^l |0\rangle.
 \ebn
 Wick's theorem allows to express this vacuum expectation value through the sum over
 all possible pairings between the linear combinations. There will
 be only three types of such pairings:
 \begin{eqnarray}
 \langle 0|\left(a+\lambda\, a^{\dag}\right)^2|0\rangle &=&
 \lambda,\label{term1}\\
 \langle 0|\Bigl(\nu(1+\mu)^{-1}a+(1+\mu)\,a^{\dag}\Bigr)^2|0\rangle
 &=& \nu,\label{term2}\\
 \langle 0|\left(a+\lambda\,
 a^{\dag}\right)\Bigl(\nu(1+\mu)^{-1}a+(1+\mu)\,a^{\dag}\Bigr)|0\rangle
 &=& 1+\mu.\label{term3}
 \end{eqnarray}

 Consider, for instance, the sum corresponding to $F_{2k,2l}$. If
 some term of this sum contains $2j$ pairings of type
 (\ref{term3}) (it is easy to understand that this number should
 be even and satisfy $0\leq j\leq\min\{k,l\}$), then it also contains $k-j$ pairings of type (\ref{term1})
 and $l-j$ pairings of type (\ref{term2}). The total number of
 such terms is equal to
 \ben
 C^{2k}_{2j}\times C^{2l}_{2j}\times (2j)!\times
 \frac{(2k-2j)!}{2^{k-j}(k-j)!}\times\frac{(2l-2j)!}{2^{l-j}(l-j)!}\,.
 \ebn
 Simplifying this combinatorial factor and taking into account the
 above remarks, we obtain a general formula for the even-even form
 factor:
 \be\label{ffeven}
 F_{2k,2l}=\sum\limits_{j=0}^{\min\{k,l\}}
 \frac{(2k)!\,(2l)!}{(2j)!\,(k-j)!\,(l-j)!}\,
 \left(\lambda/2\right)^{k-j}(1+\mu)^{2j}\left(\nu/2\right)^{l-j}.
 \eb
 Analogously, the odd-odd form factor is given by
 \be\label{ffodd}
 F_{2k+1,2l+1}=\sum\limits_{j=0}^{\min\{k,l\}}
 \frac{(2k+1)!\,(2l+1)!}{(2j+1)!\,(k-j)!\,(l-j)!}\,
 \left(\lambda/2\right)^{k-j}(1+\mu)^{2j+1}\left(\nu/2\right)^{l-j}.
 \eb
  One may also prove these results by induction,
 taking as its first step the obvious formulas
 \be
 F_{2k,0}=\langle 0|a^{2k}\hat{P}_{\lambda}|0\rangle=
 \frac{(2k)!}{k!}\left(\frac{\lambda}{2}\right)^k,\qquad
 F_{0,2l}=\langle 0|\hat{R}_{\nu}\left(a^{\dag}\right)^{2l}|0\rangle=
 \frac{(2l)!}{l!}\left(\frac{\nu}{2}\right)^l,\label{ffs}
 \eb
 and applying at the next steps the recursion relations
 \begin{eqnarray*}
 F_{k+1,l}&=&\frac{\lambda}{1+\mu}\,F_{k,l+1}+l\left(1+\mu-\frac{\lambda\nu}{1+\mu}\right)F_{k,l-1}\,,\\
 F_{k,l+1}&=&\frac{\nu}{1+\mu}\,F_{k+1,l}+k\left(1+\mu-\frac{\lambda\nu}{1+\mu}\right)F_{k-1,l}\,.
 \end{eqnarray*}

 Another problem to be considered in this section is the
 calculation of correlation functions of Bogolyubov fields in the
 hamiltonian approach. In other words, we want to compute the
 vacuum expectation values of the ordered products of
 time-dependent operators
 \be\label{btrx}
 \hat{\Okr}_{\lambda,\mu,\nu}(x)=e^{-\hat{H}x}\hat{\Okr}_{\lambda,\mu,\nu}\,e^{\hat{H}x},
 \eb
 where the operator $\hat{\Okr}_{\lambda,\mu,\nu}$ is defined by
 (\ref{btr1}). Let us start with the two-point correlation function
 $\langle\Okr_{\lambda_1,\mu_1,\nu_1}(a_1)\Okr_{\lambda_2,\mu_2,\nu_2}(a_2)\rangle$.
 Assuming that $a_2\geq a_1$ and applying the formulas
 (\ref{2pffe}) and (\ref{ffs}), one obtains
 \ben
 \langle\Okr_{\lambda_1,\mu_1,\nu_1}(a_1)\Okr_{\lambda_2,\mu_2,\nu_2}(a_2)\rangle=
 \sum\limits_{k=0}^{\infty}\;\langle0|{\Okr}_{\lambda_1,\mu_1,\nu_1}|2k\rangle\,
 \langle2k|\hat{\Okr}_{\lambda_2,\mu_2,\nu_2}|0\rangle\,e^{-2km(a_2-a_1)}=
 \ebn
 \ben
 =\sum\limits_{k=0}^{\infty}\frac{(2k)!}{(k!)^2}
 \left(\frac14\,\nu_1\lambda_2\,e^{-2m(a_2-a_1)}\right)^k
 =1+\sum\limits_{k=1}^{\infty}\frac{(2k-1)!!}{k!}
 \left(\frac12\,\nu_1\lambda_2\,e^{-2m(a_2-a_1)}\right)^k.
 \ebn
 The function characterized by such series had already appeared in
 the computation of the one-point function $\langle\Okr_V\rangle$.
 From the comparison of the last expression with the expansion
 (\ref{series}) it follows that
 \be\label{2pf}
 \langle\Okr_{\lambda_1,\mu_1,\nu_1}(a_1)\Okr_{\lambda_2,\mu_2,\nu_2}(a_2)\rangle=
 \Bigl[1-\nu_1\lambda_2\,e^{-2m(a_2-a_1)}\Bigr]^{-1/2}.
 \eb\vspace{0cm}\\
 \textbf{Remark}. In order to obtain two-point correlation
 function of fields, corresponding to $\delta$-interactions, one
 should make in (\ref{2pf}) the substitution
 \ben
 \nu_1=-\frac{V_1/2m}{1+V_1/2m},\qquad
 \lambda_2=-\frac{V_2/2m}{1+V_2/2m},
 \ebn
 and take into account the one-point functions (\ref{1point}). The
 result is then given by
 \be\label{2point}
 \langle\Okr_{V_1}(a_1)\Okr_{V_2}(a_2)\rangle=\left[
 \left(1+\frac{V_1}{2m}\right)\left(1+\frac{V_2}{2m}\right)-
 \frac{V_1V_2}{4m^2}\,e^{-2m(a_2-a_1)}\right]^{-1/2}.
 \eb
 One may easily check that it coincides with the formula
 (\ref{korr}), specialized to the case $N=2$.\vspace{0cm}\\

 Though form factor series for the multipoint correlation functions of
 Bogolyubov fields have a more complicated structure, compact expressions for
 these correlators may also be found. Note first that the form factors of the operator
 $\hat{\Okr}_{\lambda,\mu,\nu}(x)$
 are obtained from the form factors of $\hat{\Okr}_{\lambda,\mu,\nu}$ by
 the substitution
 \ben\lambda\rightarrow\lambda\,e^{-2mx},\qquad
 \nu\rightarrow\nu\,e^{2mx}.\ebn
 Next, the product
 of any two such operators is again an operator of the form
 (\ref{btr1}), multiplied by a constant. More precisely, one has
 \ben
 \hat{\Okr}_{\lambda_1,\mu_1,\nu_1}
 \hat{\Okr}_{\lambda_2,\mu_2,\nu_2}=
 c_{12}\hat{\Okr}_{\lambda_3,\mu_3,\nu_3},
 \ebn
 where the parameters $\lambda_3$, $\mu_3$ and $\nu_3$ are
 determined from the comparison of induced linear transformations,
 \ben
 \lambda_3=\lambda_1+\lambda_2\,\frac{(1+\mu_1)^2}{1-\nu_1\lambda_2},\qquad
 \mu_3=\frac{\mu_1+\mu_2+\mu_1\mu_2+\nu_1\lambda_2}{(1+\mu_1)(1+\mu_2)},\qquad
 \nu_3=\nu_2+\nu_1\,\frac{(1+\mu_2)^2}{1-\nu_1\lambda_2},
 \ebn
 and the coefficient $c_{12}$ is given by
 \ben
 c_{12}=\langle0|\hat{\Okr}_{\lambda_1,\mu_1,\nu_1}
 \hat{\Okr}_{\lambda_2,\mu_2,\nu_2}|0\rangle
 =[1-\nu_1\lambda_2]^{-1/2}.
 \ebn
 Successively applying the above observations, one may reduce any product
 of time-dependent operators to a \textit{single} operator of type
 (\ref{btr1}), multiplied by a constant factor. This constant
 clearly gives the correlation function we are looking for.

 \section{Tau-function of the Schroedinger operator \\ with point interactions}
 It was shown in Section~3 that the correlation function of
 certain Bogolyubov fields can be formally expressed by the formula
 (\ref{connection}) through the (Weinstein-Aronszajn) determinant  of the Schroedinger operator
 with $\delta$-interactions. Therefore, it is natural to assume
 that this correlator has a geometric interpretation.
 In this section, we construct a geometric invariant of
 the Schroedinger operator, simply related with the above correlation function.

 Let us choose a collection $a=(a_1,\ldots,a_N)$ of $N$ distinct points
 on the real line and suppose for
 definiteness that $a_1<a_2<\ldots<a_N$. Consider the
 $N$-parametric family of self-adjoint extensions of the operator
 $\dot{L}_a$, which correspond to separated boundary conditions
 (that is, to delta-like potentials). A general element of this
 family, $\dot{L}_{a,V}$, is given by
 \begin{eqnarray}
 \dot{L}_{a,V}=-\frac{d^2\;}{dx^2},\quad\;\;\mathrm{dom}\,\dot{L}_{a,V}=
 \{f\in H^1(\Rb)\cap H^2(\Rb\backslash
 a)\,|\,f'(a_j+0)-f'(a_j-0)=V_jf(a_j), \label{LaV} \\
 \nonumber  V_j\in\Rb,\;j=1,\ldots,N\}.
 \end{eqnarray}
 Self-adjointness of this operator implies the following important
 statement:
 \begin{prop}\label{prop1}
 If $m^2\!\in\Cb\backslash\Rb$, then the Schroedinger equation
 \be\label{scheq}(\dot{L}_{a,V}+m^2)\psi=0\eb
 has no solutions in the domain of $\dot{L}_{a,V}$.
 \end{prop}

 Let us now study the spaces of boundary values of certain \textit{local} solutions
 of the equation (\ref{scheq}). It will always be assumed that $m^2\!\in\Cb\backslash\Rb$
 and $\mathrm{Re}\,m>0$. We isolate the points $a_1,\ldots,a_N$ in the union $S=\bigcup_{j=1}^N S_j$ of
 $N$ disjoint open intervals $S_j=(x^L_j,x^R_j)$, chosen so that $a_j\in S_j$ for $j=1,\ldots,N$.
 The set $S$ is fixed once and for all, while the coordinates
 $a_1,\ldots,a_N$ are allowed to vary provided each $a_j$ stays in
 $S_j$. The whole family of operators $\dot{L}_{a,V}$ that satisfy this condition will be denoted
 by $\mathcal{L}_S$. It depends on $2N$ parameters, including the positions of delta-interactions
 $\left\{a_j\right\}_{j=1,\ldots,N}$ and their strengths $\left\{V_j\right\}_{j=1,\ldots,N}$.
 Define an auxiliary map
 \ben
 \pi:\,\mathrm{dom}\,\dot{L}_{a,V}\rightarrow
 W=\underbrace{\Cb^2\oplus\ldots\oplus\Cb^2}_{ 2n\text{
 times}}\,,
 \ebn
 \ben
 \psi\mapsto \psi^{(1)}\oplus\ldots\oplus\psi^{(N)}, \qquad
 \psi^{(i)}=\left(\begin{array}{c}\psi^{(i)}_{R,+} \\ \psi^{(i)}_{L,-}\end{array}\right)
 \oplus\left(\begin{array}{c}\psi^{(i)}_{R,-} \\
 \psi^{(i)}_{L,+}\end{array}\right),\qquad i=1,\ldots,N,
 \ebn
 where
 \ben
 \psi^{(i)}_{L,\pm}=\psi(x^L_i)\pm m^{-1}\psi'(x^L_i),\qquad\qquad
 \psi^{(i)}_{R,\pm}=\psi(x^R_i)\pm m^{-1}\psi'(x^R_i).
 \ebn

 Consider now the space of functions from the domain of $\dot{L}_{a,V}\in\mathcal{L}_S$, which solve the
 equation (\ref{scheq}) in the exterior of $S$. We will denote by
 $W^{ext}$ the image of this space in $W$ under the map $\pi$. Note that the subspace
 $W^{ext}\subset W$ is actually independent of the choice of $\{a_j\}$ and $\{V_j\}$.
 Since for any $h\in W^{ext}$ there exists a
 function $\psi\in \mathrm{dom}\,\dot{L}_{a,V}$, such that $h=\pi(\psi)$
 and $(\dot{L}_{a,V}+m^2)\psi=0$ on $\Rb\backslash \overline{S}$,
 the coordinates of $h$ should satisfy the relations
 \be\label{relh1}
 h^{(1)}_{L,-}=h^{(N)}_{R,+}=0,
 \eb
 \be\label{relh2}
 h^{(i+1)}_{L,-}=w_i\,h^{(i)}_{R,-},\qquad
 h^{(i)}_{R,+}=w_i\,h^{(i+1)}_{L,+},\qquad
 w_i=e^{-m(x^L_{i+1}-x^R_{i})},\qquad i=1,\ldots, N-1.
 \eb

 One can define along the same lines the space $W^{int}_{a,V}$ of
 boundary values of functions from the domain of $\dot{L}_{a,V}$,
 satisfying the equation (\ref{scheq}) in the interior of $S$.
 It is straightforward to check that the coordinates of any vector $g\in
 W^{int}_{a,V}$ verify
 \be\label{ni1}
 \left(\begin{array}{c}g^{(i)}_{R,-} \\
 g^{(i)}_{L,+}\end{array}\right)=
 N_i(V_i)
 \left(\begin{array}{c}g^{(i)}_{R,+} \\
 g^{(i)}_{L,-}\end{array}\right),\qquad N_i(V_i)=\left(\begin{array}{cc}
 \alpha_i(V_i) & \beta_i(V_i) \\
 \gamma_i(V_i) & \delta_i(V_i)
 \end{array}\right),\qquad i=1,\ldots,N.
 \eb
 where we have introduced  the notation
 \be\label{ni2}
 \alpha_i(V_i)=-\frac{V_i/2m}{1+V_i/2m}\,e^{-2m(x^R_i-a_i)},\qquad
 \delta_i(V_i)=-\frac{V_i/2m}{1+V_i/2m}\,e^{-2m(a_i-x^L_i)},
 \eb
 \be\label{ni3}
 \beta_i(V_i)=\gamma_i(V_i)=\frac{\;\;e^{-m(x^R_i-x^L_i)}}{1+V_i/2m}.
 \eb
 \begin{prop}\label{prop2}
 The subspaces $W^{ext}$ and $W^{int}_{a,V}$ are transverse in $W$.
 \end{prop}
 $\blacksquare$ The subspaces $W^{ext}$ and $W^{int}_{a,V}$ have zero
 intersection. In the opposite case one would be able to
 correspond to any nontrivial vector $w\in W^{ext}\cap W^{int}_{a,V}$
 a global solution of the equation (\ref{scheq}). However, the existence of such solutions
 is forbidden by the Proposition \ref{prop1}.
 Observing that $\mathrm{dim}\,W^{ext}=\mathrm{dim}\,W^{int}_{a,V}=2N$,
 one may now conclude that $W^{ext}\cup W^{int}_{a,V}=W$.
 $\square$\vspace{0.2cm}

 We can thus associate to any operator
 $\dot{L}_{a,V}\in\mathcal{L}_S$ a point $W^{int}_{a,V}$ in the
 grassmannian $Gr(2N,4N)$ of $2N$-dimensional subspaces of
 $W\simeq\Cb^{\,4N}$. Already at the present stage one
 could define the tau function as an invariant of four
 points of this grassmannian. Such construction, which can be
 thought of as a generalized cross-ratio, has been proposed in
 \cite{mason}:
 \begin{defin}
 Let W be a $2k$-dimensional complex vector space, and let
 $Gr(k,2k)$ denote the grassmannian of its $k$-dimensional
 subspaces. Given four points $W_1,W_2,W_3,W_4\in Gr(k,2k)$ in
 general position, the tau function $\tau(W_1,W_2,W_3,W_4)$ is defined as
 \ben
 \tau(W_1,W_2,W_3,W_4)=\frac{\mathrm{det}
 \left(W_1\stackrel{W_3}{\longrightarrow}W_2\right)}
 {\mathrm{det}\left(W_1\stackrel{W_4}{\longrightarrow}W_2\right)}\,,
 \ebn
 where $W_1\stackrel{W_{j}}{\longrightarrow}W_2$ denotes the
 projection of $W_1$ on $W_2$ along $W_{j}$ ($j=3,4$).
 \end{defin}

 We will give a more specialized definition of the tau function.
 First, remark that our grassmannian has a distinguished point
 $W^{int}_F\,\stackrel{def}{=}\,W^{int}_{a,0}$, corresponding to Friedrichs extension of the
 operator $\dot{L}_a$, i.~e. to ordinary second derivative
 operator on $H^2(\Rb)$. This point can be used to construct the $\mathrm{det}^*$-bundle
 over $Gr(2N,4N)$. Its fiber at the point
 $U\in Gr(2N,4N)$ is a line $\lambda^*(U)\otimes
 \lambda(W^{int}_F)$, where $\lambda(U)$ denotes maximal exterior
 power of the space $U$.

 Another distinguished point of the
 grassmannian is given by the subspace $W^{ext}$. The splitting $W=W^{ext}\oplus
 W^{int}_F$ allows to introduce the projection
 $U\stackrel{W^{ext}}{\longrightarrow}W^{int}_F$ for any $U\in
 Gr(2N,4N)$. This projection induces a linear map from
 $\lambda(U)$ to $\lambda(W^{int}_F)$, and thus defines a canonical
 section $\sigma$ of the $\mathrm{det}^*$-bundle:
 \ben
 \sigma:\,U\mapsto\mathrm{det}\left(U\stackrel{W^{ext}}{\longrightarrow}W^{int}_F\right)\in
 \lambda^*(U)\otimes \lambda(W^{int}_F).
 \ebn

 Given another map $F:\,U\rightarrow W^{int}_F$, one would be able
 to construct in analogous manner a trivializing section
 $\delta:\,U\mapsto\mathrm{det}\,F$, and to define the tau
 function
 \be\label{deftau}
 \tau(U)= \frac{\sigma(U)}{\delta(U)}=
 \frac{\mathrm{det}\left(U\stackrel{W^{ext}}{\longrightarrow}W^{int}_F\right)}{\mathrm{det}\,F}\,.
 \eb
 It is easy to see that this definition is independent of the
 choice of bases of $U$ and $W^{int}_F$. Therefore, it indeed
 gives a function on the grassmannian and, consequently, on the
 family $\mathcal{L}_S$ of Schroedinger operators.


 In order to define the map $F$, let us introduce the notion of
 auxiliary projections $P_{S_j}$ ($j=1,\ldots,N$). Suppose there
 are no delta-interactions at all and consider a single open interval
 $S'\in\Rb$. The space $W(S')=\Cb^2\oplus\Cb^2$ is
 decomposed as above into the direct sum of two-dimensional subspaces
 $W^{int}(S')$ and $W^{ext}(S')$, consisting of boundary values
 of $H^2$-solutions of Schroedinger equation without point
 interactions on $S'$ and $\Rb\backslash \overline{S'}$, correspondingly.
 We will denote by $P_{S'}$ the projection of $W(S')$ on
 $W^{int}(S')$ along $W^{ext}(S')$. The map $F:\,U\rightarrow W^{int}_{F}$ is
 now defined as the restriction to $U$ of a direct sum of such projections:
 $F=\bigl(P_{S_1}\oplus\ldots\oplus
 P_{S_N}\bigr)\Bigl|_{U}\Bigr.$. It is straightforward to check that for any vector
 $\psi\in U$ we have
 \be\label{mapf}
 (F\psi)^{(j)}=P_{S_j}\psi^{(j)}=\left(\begin{array}{c}\psi^{(j)}_{R,+} \\
 \psi^{(j)}_{L,-}\end{array}\right)\oplus N_j(0)
 \left(\begin{array}{c}\psi^{(j)}_{R,+} \\
 \psi^{(j)}_{L,-}\end{array}\right),\qquad\qquad j=1,\ldots,N.
 \eb

 Let us now obtain the explicit form of the map $F$ and of the
 canonical projection
 $P:\,U\stackrel{W^{ext}}{\longrightarrow}W^{int}_F$, putting
 $U=W^{int}_{a,V}$. First one should make some choice of coordinate bases. Remark that
 abitrary vectors $f\in W^{int}_{a,V}$ and
 $g\in W^{int}_F$ can be written (as elements of $W$) in the following way:
 \be\label{reprfg}
 f^{(j)}=\tilde{f}_j\oplus N_j(V_j)\tilde{f}_j\,,\qquad
 g^{(j)}=\tilde{g}_j\oplus N_j(0)\tilde{g}_j\,, \qquad\qquad
 j=1,\ldots,N,
 \eb
 where
 \ben
 \tilde{f}_j=\left(\begin{array}{c}f^{(j)}_{R,+} \\
 f^{(j)}_{L,-}\end{array}\right),\qquad
 \tilde{g}_j=\left(\begin{array}{c}g^{(j)}_{R,+} \\
 g^{(j)}_{L,-}\end{array}\right),
 \ebn
 and the matrices $\left\{N_j(V_j)\right\}_{j=1,\ldots,N}$ are defined by the formulas
 (\ref{ni1})--(\ref{ni3}). Hence we can represent $f$ and $g$ by
 the columns
 \ben
 f=\bigl(\tilde{f}_1^{\,T}\ldots\tilde{f}^{\,T}_N\bigr)^T,\qquad
 g=\left(\tilde{g}^{\,T}_1\ldots\tilde{g}^{\,T}_N\right)^T.
 \ebn
 It follows from (\ref{mapf})--(\ref{reprfg}) that the map $F$ is given in these coordinates
 by the identity matrix.


 In order to find the
 representation of $P$ in such coordinates, one should be able to decompose any vector
 $f\in W^{int}_{a,V}$ as $f=g+h$, with $g\in W^{int}_F$ and $h\in
 W^{ext}$. Let us obtain the relation between $f$ and $g$.
 Since the coordinates of $h=f-g$ should satisfy the conditions
 (\ref{relh1})--(\ref{relh2}), one has
 \ben
 g^{(1)}_{L,-}= f^{(1)}_{L,-},\qquad  g^{(N)}_{R,+}=
 f^{(N)}_{R,+},
 \ebn
 \ben
 g^{(j+1)}_{L,-}-w_j\, g^{(j)}_{R,-}=
 f^{(j+1)}_{L,-}-w_j\, f^{(j)}_{R,-},\qquad
 g^{(j)}_{R,+}-w_j\, g^{(j+1)}_{L,+}=
 f^{(j)}_{R,+}-w_j\, f^{(j+1)}_{L,+},\qquad
 j=1,\ldots,N-1.
 \ebn
 Using the representations (\ref{reprfg}), we may eliminate from
 these relations the extra variables $f^{(j)}_{R,-}$, $f^{(j)}_{L,+}$,
 $g^{(j)}_{R,-}$, and $g^{(j)}_{L,+}$ ($j=1,\ldots,N$). Obtained
 system of linear equations has the form
 \be\label{auxsystem}
 \left(\mathbf{1}+M_{a,V}\right)f= \left(\mathbf{1}+M_{0}\right)g,
 \eb
 where $2N\times 2N$ matrices $M_{a,V}$ and $M_0$ are defined as
 \ben
 M_{a,V}=\left(\begin{array}{ccccc}
 0        & Q_1(V_2) & 0        & .                & 0 \\
 T_1(V_1) & 0        & Q_2(V_3) & .                & 0 \\
 0        & T_2(V_2) & 0        & .                & . \\
 .        & .        & .        & .                & Q_{N-1}(V_{N}) \\
 0        & 0        & .        & T_{N-1}(V_{N-1}) & 0
 \end{array}\right),
 \ebn
 \ben
 M_{0}=\left(\begin{array}{ccccc}
 0      & Q_1(0) & 0      & .          & 0 \\
 T_1(0) & 0      & Q_2(0) & .          & 0 \\
 0      & T_2(0) & 0      & .          & . \\
 .      & .      & .      & .          & Q_{N-1}(0) \\
 0      & 0      & .      & T_{N-1}(0) & 0
 \end{array}\right).
 \ebn
 Auxiliary $2\times2$ matrices $Q_j(V_{j+1})$ and $T_j(V_{j})$,
 entering these formulas, are given by
 \ben
 Q_j(V_{j+1})=
 \left(\begin{array}{cc}
 -w_j\,\gamma_{j+1}(V_{j+1}) &  -w_j\,\delta_{j+1}(V_{j+1}) \\
 0                           &  0
 \end{array}\right),\qquad
 T_j(V_j)=
 \left(\begin{array}{cc}
 0                 & 0 \\
 -w_j\,\alpha_j(V_j) & -w_j\,\beta_j(V_j)
 \end{array}\right).
 \ebn

 The system (\ref{auxsystem}) implies that the tau function (\ref{deftau}), evaluated at the point
 $W^{int}_{a,V}\in Gr(2N,4N)$, is equal to the ratio of
 determinants
 \be\label{tau10}
 \tau(W^{int}_{a,V})=\frac{\mathrm{det}\left(\mathbf{1}+M_{a,V}\right)}
 {\mathrm{det}\left(\mathbf{1}+M_{0}\right)\;}.
 \eb
 However, this expression may be simplified.
 Since $T_j(0)Q_j(0)=0$, the matrix $\mathbf{1}+M_0$ can be represented as
 a product of a lower triangular and an upper triangular matrix with
 identities on their diagonals. Therefore, the determinant in the denominator of (\ref{tau10})
 is equal to $1$. Moreover, one can show\footnote{The proof is based on Krein's formula
 for the resolvents of self-adjoint extensions of $\dot{L}_{a}$ and lies beyond the scope of this paper.}
 that our tau-function
  does not depend on the choice of localization,
 i.~e. on the coordinates $\left\{x_j^{L,R}\right\}_{j=1,\ldots,N}$.
 In particular, we may put $x^L_j=x^R_j=a_j$ for $j=1,\ldots,N$
 and obtain the following representation:
 \be\label{taufinal}
 \tau(W^{int}_{a,V})=\mathrm{det}
 \left(\begin{array}{ccccc}
 \mathbf{1}  & \tilde{Q}_1 & 0           & .               & 0 \\
 \tilde{T}_1 & \mathbf{1}  & \tilde{Q}_2 & .               & 0 \\
 0           & \tilde{T}_2 & \mathbf{1}  & .               & . \\
 .           & .           & .           & .               & \tilde{Q}_{N-1} \\
 0           & 0           & .           & \tilde{T}_{N-1} & \mathbf{1}
 \end{array}\right),
 \eb
 where
 \ben
 \tilde{Q}_j=\frac{e^{-m(a_{j+1}-a_j)}}{1+\frac{V_{j+1}}{2m}}
 \left(\begin{array}{cc}
 -1 & \frac{V_{j+1}}{2m} \\ 0 & 0
 \end{array}\right),\qquad
 \tilde{T}_j=\frac{e^{-m(a_{j+1}-a_j)}}{1+\frac{V_{j}}{2m}}
 \left(\begin{array}{cr}
 0 & 0 \\ \frac{V_{j}}{2m} & -1
 \end{array}\right),\qquad j=1,\ldots,N-1.
 \ebn

 Finally, it is worth mentioning that the tau function $\tau(W^{int}_{a,V})$
 and the correlation functions of Bogolyubov fields, associated to $\delta$-interactions, are
 related by
 \be\label{fin}
 \tau(W^{int}_{a,V})=
 \left[\frac{\langle\Okr_{V_1}(a_1)\ldots\Okr_{V_N}(a_N)\rangle}
 {\langle\Okr_{V_1}\rangle\ldots\langle\Okr_{V_N}\rangle}\right]^{-2}.
 \eb
 This formula may be checked explicitly for small values of $N$
 by comparison of (\ref{taufinal}) and
 (\ref{korr}).\vspace{0.2cm}\\
 \textbf{Example}. For the two-point tau function one has
 \ben \tau(W^{int}_{a,V})=
 \mathrm{det}\left(\begin{array}{cc}\mathbf{1} & \tilde{Q}_1 \\
 \tilde{T}_1 & \mathbf{1}\end{array}\right)=
 \mathrm{det}\left(\mathbf{1}-\tilde{T}_1\,\tilde{Q}_1\right)=
 1-\frac{V_1/2m}{1+V_1/2m}\frac{V_2/2m}{1+V_2/2m}\,e^{-2m(a_2-a_1)}\;. \ebn
 Comparing this expression with the formulas (\ref{1point}) and
 (\ref{2point}), we may check the validity of (\ref{fin}) for
 $N=2$.

 \section{Discussion}
 It is well-known that in order to describe delta-interactions in
 two and three dimensions, one needs to renormalize the strengths
 of point sources. This obstacle complicates the construction
 of fields, associated to delta-interactions in higher dimensions.
 In particular, a naive generalization,
 $\hat{\Okr}_V(t,\vec{x})=\exp\left\{-\frac12\, V\hat{\varphi}^2(t,\vec{x})\right\} $, does not work,
 since even the vacuum expectation value of such an operator is
 infinite. It would be interesting to understand how the
 renormalization can be described (if it indeed can) in terms of
 Bogolyubov fields.

 It would also be interesting to consider instead of the Schroedinger
 operator the hamiltonian of Dirac fermions in $2+1$ dimensions in
 the background of magnetic vortices. Its resolvent is the principal ingredient
 in the computation of induced fermionic vacuum quantum numbers. It also contains the
 information about the scattering of fermions, their bound states,
 etc. It seems that this resolvent may be expressed along the lines of the
 present paper (at least, for
 some values of self-adjoint extension parameters) through the
 ratio of certain correlation functions in the two-dimensional
 euclidean quantum field theory, described by the Dirac action
 with a parity-breaking term. Thus one needs to find form factors and correlation
 functions of the Bogolyubov fields, associated to magnetic vortices.
 The absence of parity-breaking term means that the resolvent is calculated at zero energy.
 Corresponding Bogolyubov fields reduce in this case to the exponential fields of the
 $\text{SG}_{ff}$-theory.

 Yet another area of applications of Bogolyubov fields lies in
 quantum optics, where they correspond to the so-called
 ``squeezed'' operators \cite{optics}, describing the
 minimum-uncertainty states.

 \appendix
 \section{Gaussian integration}
 Suppose we are given:
 \begin{itemize}
 \item two real vectors $\varphi=(\varphi_1,\ldots,\varphi_M)^T$ and $\mu=(\mu_1,\ldots,\mu_N)^T$;
 \item a vector $J=(J_1,\ldots,J_M)^T$;
 \item two real invertible symmetric square matrices $A$ and $B$ of
 the corresponding sizes $M\times M$ and $N\times N$;
 \item an $M\times N$ matrix $C$.
 \end{itemize}
 If we denote $\mathcal{D}\varphi=d\varphi_1\ldots d\varphi_M$,
 $\mathcal{D}\mu=d\mu_1\ldots d\mu_N$,
 then the following formulas hold:
 \begin{eqnarray}
 \nonumber &\;&Z_A = \int\mathcal{D}\varphi\; e^{-\frac12\, \varphi^T\!
 A\varphi} =
 \frac{\;\;\;\left(2\pi\right)^{M/2}}{\sqrt{\mathrm{det}\,A}},\\
 \nonumber &\;& Z_A^{-1}\int\mathcal{D}\varphi\; e^{-\frac12\, \varphi^T\!
 A\varphi+J^T\varphi} = e^{\;\frac12\, J^T\! A^{-1}J}, \\
 \nonumber &\;& Z_A^{-1}\int\mathcal{D}\varphi\; \varphi_i\varphi_j\; e^{-\frac12\, \varphi^T\!
 A\varphi} = A^{-1}_{\;ij},\\
 \nonumber &\;& Z_A^{-1}\int\mathcal{D}\varphi\; \varphi_i\varphi_j\;e^{-\frac12\, \varphi^T\!
 A\varphi+J^T\varphi} = \left[A^{-1}_{\;ij}+
 \left(A^{-1}J\right)_i\left(A^{-1}J\right)_j\right]e^{\;\frac12\, J^T\!
 A^{-1}J},
 \end{eqnarray}
 \begin{eqnarray}
 \nonumber Z_{A,B,C}=\int\mathcal{D}\varphi\,\mathcal{D}\mu \;
 e^{-\frac12\, \varphi^T\! A\varphi-\frac12\, \mu^T\! B\mu-\varphi^T\!
 C\mu}={\left(2\pi\right)^{\frac{M+N}{2}}}\biggl/{\sqrt{\mathrm{det}\left(
 \begin{array}{ll} A & C \\ C^T & B \end{array}\right)}}=
 &\;& \\
 =\left(2\pi\right)^{\frac{M+N}{2}}\bigl/{\sqrt{\mathrm{det}\,A\;\mathrm{det}\left(B-C^T A^{-1}C
 \right)}}=&\;&\label{ff}
 \\
 \nonumber ={\left(2\pi\right)^{\frac{M+N}{2}}}\bigl/{\sqrt{\mathrm{det}\,B\;\mathrm{det}\left(A-C
 B^{-1}C^T \right)}}\,,\;\;\,&\;&
 \end{eqnarray}
 \begin{eqnarray}
 \nonumber Z_{A,B,C}^{\;\;-1}\int\mathcal{D}\varphi\,\mathcal{D}\mu \;
 \varphi_i\varphi_j\,e^{-\frac12\, \varphi^T\! A\varphi-\frac12\, \mu^T\! B\mu-\varphi^T\!
 C\mu}= \left(A-CB^{-1}C^T\right)^{-1}_{ij}= &\;&\\
  = A^{-1}_{\; ij}+\left[A^{-1}C\left(B-C^T A^{-1}C\right)^{-1}\!C^T
 A^{-1}\right]_{ij}.& \;&\label{krein1}
 \end{eqnarray}

 Now let the vectors $\varphi$ and $\mu$ be complex, and let the
 matrices $A$ and $B$ be hermitian. Then, introducing the measures
 \ben
 \mathcal{D}\varphi\mathcal{D}\bar{\varphi}=2^M \prod_{j=1}^M d(\mathrm{Re}\,\varphi_j)\,d(\mathrm{Im}\,\varphi_j),
 \qquad\qquad
 \mathcal{D}\mu\mathcal{D}\bar{\mu}=2^N \prod_{j=1}^N d(\mathrm{Re}\,\mu_j)\,d(\mathrm{Im}\,\mu_j),
 \ebn
 one may obtain the analogs of the above relations:
  \begin{eqnarray}
 \nonumber &\;&Z_A = \int\mathcal{D}\varphi\,\mathcal{D}\bar{\varphi}\;
 e^{-\varphi^{\dag}\! A\varphi}
 = \frac{\;\;\left(2\pi\right)^{M}}{\mathrm{det}\,A},\\
 \nonumber &\;& Z_A^{-1}\int\mathcal{D}\varphi\,\mathcal{D}\bar{\varphi}\;
 e^{-\varphi^{\dag}\! A\varphi+J^{\dag}\varphi+\varphi^{\dag}J}
 = e^{\; J^{\dag}\! A^{-1}J},\\
 \nonumber &\;& Z_A^{-1}\int\mathcal{D}\varphi\,\mathcal{D}\bar{\varphi}\;
 \varphi_i\bar{\varphi}_j\; e^{-\varphi^{\dag}\! A\varphi} = A^{-1}_{\;ij},\\
 \nonumber &\;& Z_A^{-1}\int\mathcal{D}\varphi\,\mathcal{D}\bar{\varphi}\;
 \varphi_i\bar{\varphi}_j\;e^{- \varphi^{\dag}\! A\varphi+J^{\dag}\varphi+\varphi^{\dag}J}
 = \left[A^{-1}_{\;ij}+ \left(A^{-1}J\right)_i\left(\overline{A^{-1}J}\right)_j\right]
 e^{\; J^{\dag}\! A^{-1}J},
 \end{eqnarray}
 \begin{eqnarray}
 \nonumber Z_{A,B,C} =\int\mathcal{D}\varphi\,\mathcal{D}\bar{\varphi}\;\mathcal{D}\mu\,\mathcal{D}\bar{\mu} \;
 e^{- \varphi^{\dag}\! A\varphi-\mu^{\dag}\! B\mu-\varphi^{\dag} C\mu-\mu^{\dag}C^{\dag}\varphi}
 ={\left(2\pi\right)^{M+N}}\bigl/{\mathrm{det}\left(
 \begin{array}{ll} A & C \\ C^{\dag} & B
 \end{array}\right)}=
 \\
 ={\left(2\pi\right)^{M+N}}\bigl/\left[\mathrm{det}\,A\;\mathrm{det}\left(B-C^{\dag}A^{-1}C \right)\right]=
 {\left(2\pi\right)^{M+N}}\bigl/\left[\mathrm{det}\,B\;\mathrm{det}\left(A-C
 B^{-1}C^{\dag} \right)\right],
 \end{eqnarray}
 \begin{eqnarray}
 \nonumber Z_{A,B,C}^{\;\;-1}
 \int\mathcal{D}\varphi\,\mathcal{D}\bar{\varphi}\;\mathcal{D}\mu\,\mathcal{D}\bar{\mu}\;
 \varphi_i\bar{\varphi}_j\,e^{-\varphi^{\dag}\! A\varphi- \mu^{\dag}\!
 B\mu-\varphi^{\dag}C\mu-\mu^{\dag}C^{\dag}\varphi}= \\
 =\left(A-CB^{-1}C^{\dag}\right)^{-1}_{ij}=
 A^{-1}_{\; ij}+\left[A^{-1}C\left(B-C^{\dag}A^{-1}C\right)^{-1}\!C^{\dag} A^{-1}\right]_{ij}.
 \end{eqnarray}

\end{document}